# Resistive switching and Schottky barrier modulation at CoPt/ Ferroelectric-like MgZnO interface for non-volatile memories

Mohamed Belmoubarik [1,2*], Muftah Al-Mahdawi [2], George Machado Jr. [3], Tomohiro Nozaki [2] Cláudia Coelho [3] Masashi Sahashi [2] and Weng Kung Peng [3,4]

[1] Institute of Applied Physics, Mohammed VI Polytechnic University, Lot 660, Hay Moulay Rachid Ben Guerir, 43150, Morocco

[2] Department of Electronic Engineering, Tohoku University, Sendai 890-8579, Japan

[3] International Iberian Nanotechnology Laboratory, INL, Av. Mestre José Veiga s/n, Braga, Portugal

[4] Songshan Lake Materials Laboratory, 523-808, Dongguan, China

*Corresponding authors      mohamed.belmoubarik@um6p.ma

## Abstract

Ferroelectric memristors have attracted much attention as a type of nonvolatile resistance switching memories in neuromorphic computing, image recognition, and information storage. Their resistance switching mechanisms have been studied several times in perovskite and complicated materials systems. It was interpreted as the modulation of carrier transport by polarization control over Schottky barriers. Here, we experimentally report the isothermal resistive switching across a CoPt/MgZnO Schottky barrier using a simple binary semiconductor. The crystal and texture properties showed high-quality and single-crystal $Co_{0.30}Pt_{0.70}/Mg_{0.20}Zn_{0.80}O$ hetero-junctions. The resistive switching was examined by an electric-field cooling method that exhibited a ferroelectric Curie temperature of MgZnO close to the bulk value. The resistive switching across CoPt/MgZnO Schottky barrier was accompanied by a change in the Schottky barrier height of 26.5 meV due to an interfacial charge increase and/or orbital hybridization induced partial reversal of the MgZnO polarization. The magnitude of the reversed polarization was estimated to be a reasonable value of 3.0 (8.25) $\mu C/cm^2$ at 300 K (2 K). These findings demonstrated the utilities of CoPt/MgZnO interface as a potential candidate for ferroelectric memristors and advanced spintronics applications.                                            (180 words)

# 1) Introduction

Polarization reversal is a special physical effect in ferroelectric (FE) crystals that gave birth to new electronic functionalities such as the nonvolatile information storage and surface charge modulation. The Integration of these properties with CMOS devices has been achieved, but the expected technological impact remains unreachable.[1,2] With regards to conventional perovskite ferroelectrics, hardship with scaling down the FE barrier below ~100 nm, necessity of implementing oxides ($SrTiO_2$, LSMO and $SrRuO_3$) as bottom electrodes to obtain high-quality crystals and incompatible windows for crystal synthesis are the main features blocking a smooth integration.[3,4] Recently, it has been clear that exploiting the full capabilities of highly scaled and integrated ferroelectrics stand in need of a different material set, of course, beyond the perovskite titanate palette. To do so, it is important to explore structurally simpler polar crystals and analyze the possibility of a reversable electric dipoles by an electric field. Suggestions toward this approach are present in early literature, for example, in 1952 by Von Hippel,[5] and in 1957 by Megaw;[6] however, vast studies that tackle these suggestions have not been reported. Two novel ferroelectric materials were under the scope during the past decade with extensive investigations: $HfO_2$ and AlScN.[1,7] In $HfO_2$, crystal properties such grain size, strain, and defect chemistry can result into a stable and switchable thin film, while in AlN, substituting ~20% Sc for Al will achieve the same. These findings orient the research community to a rich new source of FE materials, where polar crystals are strong candidates as ferroelectric hosts. interestingly, two-dimensional materials attracted much interest as potential candidates due do its piezoelectric and FE properties[8] and the realization of high Curie temperature ($T_C$) FE thin films (~100–300 nm) such as $CuCrS_2$[9] and α-$In_2Se_3$.[10,11]

Zinc oxide (ZnO) – based material is widely used semiconductor in electronic and optoelectronic devices, especially for short-wavelength lights generation and detection,[12,13] due to its wide band-gap of 3.34 eV with a Wurtzite (*wz*) crystal, high exciton binding energy at room temperature (RT), higher quantum efficiency, and amenability to wet chemical etching.[14–16] Changing the ZnO material properties by means of metal doping induces versatile physical properties, such electric, magnetic and ferroelectric,[17,18] and making of it an attractive tool for the fabrication of low power electronic devices.[19,20] Onodera *et al.* reported ferroelectricity in bulk $Zn_{1-x}B_xO$ (B = Li, Mg), but the polarization values were unusually small (0.06–0.9 µC/cm$^2$).[21] However, defect-mediated pseudoferroelectricity can emerge in non-doped ZnO thin films[22], and also in doped ZnO by Mg, Li, Cd, V etc [21,23–28]. The ferroelectricity of *wz*-(Mg,Zn)O alloys was theoretically studied by several groups and supported the presence of FE property in wurtzite-like structures[29–34]. Tagantsev *et. al*[29] categorized doped-ZnO as a pseudoferroelectric material based on the Landau theory. In addition, Herng *et al.*[22] have experimentally reported pseudoferroelectricity in ZnO by a combination with first-principles calculations, and they have proposed an oxygen vacancy-assisted pseudoferroelectricity for the controversial (Mg,Zn)O materials[34]. The weak insulating property of these materials screens the applied electric field through conduction electrons and reduces their ferroelectricity. Because their breakdown field ($E_{bd}$) is significantly smaller than their coercive field ($E_c$) and piezoelectric properties[35], the crystal strain in *wz*-(Mg,Zn)O is easily induced by introducing an electric field. Also, the switching barrier energy ($E_{sb}$) of *wz*-(Mg,Zn)O stays higher (in the range of 0.3–0.6 eV/ f.u.)[31,33,36] compared to other FE materials[35]. Despite having such low $E_{bd}$ and high $E_{sb}$ and $E_C$ values, *wz*-(Mg,Zn)O crystals have been found to be ferroelectric in few studies.[17,21,24,37] More recently, Ferri *et al.*[18] reported a giant switchable polarization of >100 µC/cm$^2$ in an optimized 500-nm-thick *wz*-MgZnO films (Mg = 30–37%) deposited on Pt electrodes, at coercive fields within 3–4 MV/cm at RT, supporting the earlier expectations about polar oxides.[33] Interestingly, the FE MgZnO-based non-volatile resistive memory presents an opportunity for adopting a two-terminal cross-bar device architecture, enabling scalability beyond traditional silicon CMOS. Memory elements with top-to-bottom contacts can fall into various memristive systems, including magnetic random-access memory, filamentary resistive-switch, phase change memory, and ferroelectric memory.[35] Among these, the ferroelectric memory device driven by spontaneous polarization stands out due to its potential for high-speed (read time < 55 ns), multi-bit feasibility (2-4 bits), and low-power nonvolatile memory operations (<10 fJ/bit).[3,35] Very recently, robust switchable polarization by an external voltage and a coupled giant modulation of the electronic transport behavior of Mg- and Cr-doped ZnO was reported (high ON/OFF ratio ≈ $10^4$)[38,39].

On the other hand, Schottky contacts (SCs) on (Mg,Zn)O alloys are important and useful in many applications, including high-electron-mobility transistors,[40,41] metal semiconductor field-effect transistors,[42] ultraviolet photodetectors,[43,44] gas sensors,[45,46] and piezoelectric nanogenerators.[47,48] SCs on n-type (Mg,Zn)O, are difficult to be prepared compared to ohmic contacts because of the high donor concentration at the n-ZnO surface due to native defects, such as oxygen vacancies and/or zinc interstitial atoms. Hence, the main hurdle of maintaining a high-quality MgZnO SC on a metallic electrode while generating and exploiting its FE behavior. Here, the SC with a FE MgZnO will serve as a demonstration of ferroelectric memristors and polarization reversal-induced barrier height change in MgZnO SCs, as exhibited in the ferroelectric Schottky diodes.[49] In this case, the potential profile of the Schottky barrier, formed between the metal and the FE semiconductor, can be reversibly changed through the flipping of FE polarization. On the other hand, *wz*-(Mg,Zn)O alloy has been proposed as a viable tunnel-barrier option for low-power magnetic tunnel junctions (MTJs).[50,51] Previous studies have highlighted the CoPt/ *wz*-MgZnO(0001) interface as a versatile component for spintronics devices[17,52] due to the crystallographic compatibility of *wz*-MgZnO with the fcc-type CoPt alloys that exhibit high perpendicular magnetic anisotropy and high ferromagnetic (FM) $T_C$ [53]. This will permit the implementation of the FE-MgZnO as a tunneling barrier inside an MTJ and then the realization of an electric polarization modulation of the magnetoresistance, hence, a multi-state memory.[2] Also, the FM-CoPt/FE-MgZnO combination will also permit the evaluation of the charge density change at the CoPt interface by FE polarization reversal, and therefore the change in their magnetic anisotropy[52]. In addition, the CoPt/(Mg,Zn)O interface could offer potential applications in spin injection, LED technology, and spin-photonics devices, as exemplified in limited semiconductor configurations such as Co-Pt/p-Si and CoPt/GaAs.[54,55] However, to the best of our knowledge, evidence and investigation of ferroelectricity of MgZnO and/or MgZnO SCs on FM electrodes, which is compatible with spintronics applications, at the level of a single nanocrystallite have thus far been lacking considering that in most of the cases FE-MgZnO thin films were obtained using the metallic Pt electrodes[12,18,38,56].

In this report, we demonstrate the epitaxial growth of high-quality CoPt/MgZnO SC and verified the ferroelectricity behavior of a 100-nm-thick *wz*-MgZnO thin film by the electric field cooling method as adopted in our previous report[17], and consecutive DC double-sweep voltage ramps. The extraction and analysis of the ferroelectricity driven Schottky barrier height change during an isothermal resistive switching was performed along with a comparison of the reversed FE polarization with previous theoretical values. The present results will open many new possibilities for implementing a non-perovskite FE *wz*-MgZnO thin films into electric devices and achieving hybrid interfaces with FM electrodes for advanced spintronics applications.

## 2) Experimental methods

Samples of this report have the following structure: c-plane $Al_2O_3$ substrate/ Pt(30, 700)/ $Co_{0.30}Pt_{0.70}$ (10, 500)/ $Mg_{0.08}Zn_{0.92}O$(100, 500)/ Cu(600, 25) [stack-A], c-plane $Al_2O_3$ substrate/ Pt(30, 700)/ $Co_{0.30}Pt_{0.70}$ (10,

500)/ Mg$_{0.20}$Zn$_{0.80}$O(220, 400)/ Cu(600, 25) [stack-B], where the numbers in parentheses are layer thicknesses in nm and growth temperature in °C, respectively. The Al$_2$O$_3$ substrate was thermally annealed at 700 °C for 10 min to enhance the flatness of the surface. The metallic electrodes of Pt and Co$_{0.30}$Pt$_{0.70}$ were grown at a deposition rate of 1.2–1.8 nm/min, respectively, using a magnetron co-sputtering system. Then, MgZnO thin films were deposited by MBE technique at 500 °C using an optimized O$_2$ flow of 2.0 SCCM (cubic centimeter per minute at STP) to maintain the magnetic properties of Co$_{0.30}$Pt$_{0.70}$ electrodes. The MBE chamber is accompanied with Zn and Mg Knudsen cells of high purity materials (4N) and attached to the co-sputtering chamber with an ultra-vacuum channel. The deposition rate of MgZnO thin films was in the range of 4.7–5.6 nm/min. The top Cu electrode was deposited at RT to avoid its oxidation. The surface treatment of the Co$_{0.30}$Pt$_{0.70}$ was omitted to maintain the crystalline order and magnetic properties. We note that the choice of Pt/Co$_{0.30}$Pt$_{0.70}$ as a bottom buffer layer is based on the reasons elaborated in our previous report.[50] The ex-situ surface morphologies, crystalline properties, composition ratio of both MgZnO and Co$_{0.30}$Pt$_{0.70}$ alloys, and magnetic properties of these stacks were investigated by atomic force microscopy (AFM), in-plane and out-of-plane X-ray diffractometer (XRD) (Bruker D8 Discover), X-ray fluorescence (XRF) and vibrating-sample magnetometer (VSM), respectively. The heterostructures crystalline quality was characterized using high-resolution transmission electron microscopy (HR-TEM). After deposition, the samples were patterned into a current-perpendicular-to-plan (CPP) circular devices of 200-μm diameters using dry etching, passivation of SiO$_2$ insulation and lift-off techniques (see **Fig.4 (a)**). The microfabrication was designed to minimize the effect of ZnO side milling and its degradation. The back-to-back Schottky diode (BBSD) diodes were characterized by the conventional two-probes current–voltage (*I–V*) measurements where the voltage bias was applied to the top electrode using an impedance analyzer and DC current measurement. From here we refer to $Co_{0.30}Pt_{0.70}$ and $wz$-$Mg_xZn_{1-x}O$ materials, where x is the Mg composition, as CoPt and MgZnO for simplicity.

## 3) Structural and texture properties.

The crystalline properties of stack-A are shown in **Figure 1 (a)**. In the out-of-plane ω–2θ-scans, around Pt(111) and $Co_{0.30}Pt_{0.70}$(111) diffraction peak satellites due to Laue oscillations were observed, thus the achievement of coherent growth and high-quality thin films with a small interface roughness. $wz$-ZnO(0002) and (0004) peaks were sharp without any additional phases. The other small peaks were confirmed to result from Cu-$K_β$ or W-$K_{α,β}$ radiations. Also, as shown in **Figure 1 (a)**, the in-plane 2θχ-φ scans around asymmetric plans revealed the presence of single-phase of $Al_2O_3$, CoPt(Pt) and ZnO. In addition, the spectra of the in-plane φ-scans around asymmetric reflections of stack-A materials (not shown here) exhibited a six-fold in-plane symmetry with sharp peaks for all layers with a clear epitaxial relationship: $Al_2O_3$(0001)[30–30]/ Pt(111)[220]/ $Co_{0.30}Pt_{0.70}$(111)[220]/ MgZnO(0001)[11–20]. The magnetic properties of the bottom Pt/$Co_{0.30}Pt_{0.70}$ electrode didn't change after the deposition of the MgZnO thin films. Hence, the MBE technique allowed high-quality and single-crystal $wz$-ZnO films deposition on an FM Pt/$Co_{0.30}Pt_{0.70}$ buffer at a moderate temperature of 500 °C. For the selected bottom electrode of Pt/CoPt, the crystal phase-shift of MgZnO from wurtzite to cubic occurs at an Mg content of 37-40% similar to previous reports.[12,20]

The *ex-situ* surface morphologies of $Co_{0.30}Pt_{0.70}$ and MgZnO films are illustrated in **Fig. 1 (b)** and **(c)**. The CoPt surface, with a roughness average (Ra) of 0.26 nm *vs.* 0.18 nm for the flat sapphire substrate, exhibited a flat surface due to the small lattice mismatch between $Al_2O_3$(0001) and Pt(111) estimated to be 0.86%.[50] The top-side AFM images of 100-nm-thick MgZnO film is flat with an Ra of 2.45 nm, as shown in **Figure 1 (c)** consistent with streaky patterns of the reflection high-energy electron diffraction (RHEED) (not shown here) and the small lattice mismatch between ZnO(0001) and CoPt(111) estimated to be 4.8%. HR-TEM images of a stack-A along $Al_2O_3$ [0001] axis is shown in **Figure 1 (d)**. For all the composing layers were clearly distinguished and exhibited good crystallinity with abrupt interfaces. The absence of any element segregation or out-diffusion is confirmed within the enlarged green rectangle of **Figure 1 (e)** with some imperfections at the interface. In the latter one, the white rectangles corresponding to ZnO and $Co_{0.30}Pt_{0.70}$ films exhibited clear atomic alignments of both hcp-(0001) and fcc-(111), respectively. The blue and red circles inside the ZnO area exhibited the hcp classical zigzag pattern of Zn and O atoms as assigned by the violet line, while the black and gray circle inside CoPt area represented Co and Pt atoms of a well-textured $CoPt_3$-like alloy. Consequently, RHEED, AFM and HR-TEM images confirmed the deposition of high-quality Pt-based FM electrode and MgZnO thin film using a combination of co-sputtering and MBE deposition techniques and post-thermal treatments.

## 4) Ferroelectricity-like in *wz*-MgZnO deposited on CoPt electrode

The ferroelectricity of *wz*-MgZnO alloys was extensively studied theoretically[29,30,33,34,57] and experimentally[17,18,21] to conclude the possibility of its realization. Most notably for tunnel barrier fabrication, Mg doping lowers the free carriers concentration and enhances insulating property.[58] In our previous report, we could demonstrate the FE property of a 7nm-thick $Mg_{0.23}Zn_{0.77}O$ tunneling barriers with a $T_C$ of 316 K; which is close to the bulk value of 320-350 K;[38] by investigating the tunneling electroresistance (ER) effect[59] using the electric field cooling (EFC) method. The FE effect could be explained by the shift of CoPt surface potential and the change of the average barrier heights due to the partial reversal of MgZnO electric polarization ($P_r$).[17,52] The schematic of the device and the effect of the

±EFC process on the polarization alignment is shown in **Figure 2 (a)**. The direct characterization of FE property of *wz*-MgZnO is mostly performed by the piezoresponse force microscopy (PFM), and high $E_C$ within 500-3000 kV/cm (see **Table S1**), versus lower orders within 5-50 kV/cm for other FE materials (such as BaTiO$_3$, PbTiO$_3$ and PZT[35,60]), is needed to reverse the electric polarization. Along with the high $E_{sb}$ as presented in the introduction, these facts made difficult the direct observation of the polarization reversal by PFM in *wz*-MgZnO deposited on CoPt buffer layer due to the unavoidable oxygen vacancies-induced leakage current (see **Figure S2**). We observed an increase of the reversed polarization of *wz*-MgZnO at negative electric fields before reaching the limit for current leakage, which is a good indication of pseudoferroelectricity. Therefore, the EFC method from above $T_C$ of MgZnO was adopted to align the polarization in both up and down directions (**Figure 2 (a)**). The reversal of $P_r$ results in a difference of the equilibrium surface charge at the CoPt interfaces (**Figure 2 (b)**), and the charge difference is negative (positive) at the CoPt surface after +EFC (−EFC), due to the formation of a positive $P^+$ (negative $P^-$) state [**Figure. 2 (a)** and **(b)**]. Consequently, a high (low) Schottky barrier height (SBH) is expected to be formed at the CoPt/MgZnO interface of the dominant SC resulting into a resistive switching, as mentioned in **Figure 2 (b)**.

In order to achieve the EFC on a MgZnO-SC without using a large electrical field, the Mg content, thickness of MgZnO barrier, applied voltage were selected to be 8%, 44 nm and ±1.5 V, respectively, which permit the measurement of high resistances at a voltage bias of +10 mV without breaking the device. The CPP device was cooled from a temperature of 300 K higher than the FE phase-transition temperature ($T_C$ = 260−280 K) of the bulk Mg$_{0.08}$Zn$_{0.92}$O[21] down to a much lower value of 4 K. It is worth mentioning that a magnetic field of 10 kOe was always applied parallel to the sample plane in order to align the magnetization directions of the CoPt electrode, and consequently cancel any tunneling magnetoresistance (TMR) effect. Afterwards, the temperature-dependent device resistance after each EFC process was measured at a low bias of 10 mV during heating up to 300 K (zero-field heating, ZFH) with a rate of +2 K/min, as plotted in **Figure 2 (c)**. Also, the after-mentioned experimental results were qualitatively similar among pillars from the same substrate but with different magnitudes.

The temperature-dependent ZFH resistance ($R - T$) curves after EFC at 0 V and +1.5 V are shown in **Fig. 2 (c)**. The resistance after −1.5 V EFC is larger than that after +1.5 V EFC, thus, the resistive switching of the fabricated CoPt/MgZnO/Cu BBSD was demonstrated by means of the EFC method. Since the conduction through MgZnO-based BBSD is almost dominated by the thermionic effect, the ER ratio can be defined as: $ER = J_{\text{LRS}}/J_{\text{HRS}}$, where $J_{\text{LRS}}$ and $J_{\text{HRS}}$ are the current densities for the low resistance state (LRS) and high resistance state (HRS), respectively, at a low bias-voltage. The *ER* ratio was 240% at low temperatures and increased up to 400 % in other pillars. Notably, the electro-migration of oxygen vacancies and defects[61,62] and/or conductive filament formation[63,64] can be ruled out as the origin of the observed ER effect. This is because the ER ratio largely vanished at temperatures above 300 K (inset of **Figure 2(c)**), whereas an electromigration or filament-based resistive switching should remain at elevated temperatures near 400 K.[64–67] Notably, the coexistence of TMR effect for both LRS and HRS using the CoPt/MgZnO (7 nm) interface (not shown here), deposited with the same conditions of stack-B, rules out the non-ferroelectric origin of the resistive switching since TMR is very sensitive to the interfacial magnetic spin polarization[68] that will diminish along with the magnetic properties of CoPt by the accumulation of oxygen vacancies or filaments formation[69,70]. Consequently, the observed resistive switching in the epitaxial CoPt/*wz*-MgZnO SC is likely occurring due to the partial polarization reversal due to vacancies-mediated pseudoferroelectricity within a homogeneous MgZnO thin film.

The epitaxial conditions of the oxygen-rich environment and the surface stability of the O-polar face compared to the Zn-polar one[71] will favor the O-polar growth corresponding to $P^+$ state. Therefore, the polarization of the MgZnO film grown on the fcc-CoPt is expected to stabilize in a downward ($P^+$) state. As shown in the HR-TEM image of a virgin film near the CoPt/MgZnO interfaces, a zig-zag pattern that is consistent with the O-polar growth was observed (**Figure 1 (e)**). Furthermore, *R–T* curves after EFC at 0 V and +1.5 V were similar, reflecting the same stable $P^+$ and LRS (**Figure 2(b)**). After EFC at −1.5 V, the polarization reversed to the upward direction ($P^-$) resulting in an accumulation of negative charges at the CoPt interface, thus, an increased SBH and a HRS is generated (**Figure 2 (b)**). Hence, the resistive switching induced by the EFC method can be explained by a ferroelectricity-like induced SBH modulation due to a polarization reversal at the CoPt/MgZnO/Cu interfaces, similar to previous metal/FE Schottky barriers.[72,73] In order to sustain this explanation, we plotted the temperature dependence of ln(*ER*) (inset of **Figure 2 (c)**) which is proportional to both $P_r$ the reversed polarization and polarization-reversal induced SBH change $\Delta\Phi_B$, $ER = exp(\Delta\Phi_B/k_BT)$ as it will be elaborated later. This dependency follows a power-rule with a $T_C$ of 239 ± 5 K and a β coefficient of 0.86, which indicates a non-uniform polarization reversal as observed in the FE SrTiO$_3$.[74] The estimated $T_C$ is close to the bulk value (250 K) determined from the temperature-dependent dielectric constant.[21] A quantitative estimation of $P_r$ and *ER* using the experimentally extracted SBHs at LRS and HRS states will be introduced at section 6). We should note that the non-zero *ER* ratio above the estimated $T_C$ of MgZnO (inset of **Figure 2c**) is thought to be coming from a non-dominant oxygen-vacancies based resistive switching.

    The FE-type resistive switching was investigated again by consecutively sweeping the voltage several cycles at ±1, ±2, ±3, ±4 V for a pillar of stack-B at RT (**Figure 3**). The curve for ±1 V does not show any hysteresis because the voltage range is insufficient to induce ferroelectric switching. For voltage of ±2 V or more, hysteretic resistance switching is realized. A gradual increase of the resistance after each negative

voltage ramping corresponds to the partial poling of to the $P^-$ state. On the other hand, applying positive voltages tends to bring the MgZnO polarization to the $P^+$ state, the ground low state. The ER ratio measured at -1 V as a function of sweeping maximum voltage exhibited a saturation behavior corresponding to a ferroelectric memristor since the *ln*(ER ratio) is proportional to the reversed electric polarization (see **Eq (5)** and **(6)** in section 6)). Here, we should mention that the set to the high state ($P^-$) and full poling of MgZnO is limited by the oxygen vacancies that reduces the order of reversed electric polarization $P_r$ similar to our previous report for ultrathin films at the tunneling regime[17].

## 5) Extraction of the MgZnO Schottky barrier heights.

Challenging the realization of an isothermal resistive switching was performed for stack B, where the Mg content is 20% and $T_C$ is expected to exceed RT (330 K for bulk-based experiment[21] vs. 390 K for theoretical reports[75]). The MgZnO was selected to be 220 nm to suppress the tunneling effect and decrease the leakage current of MgZnO thin film epitaxially deposited at 500 °C. The current-voltage (*I–V*) characteristics for a DC voltage swept as follows 0 → +4 V → −4 V → 0 V on a CPP device of stack-B (**Figure 4 (a)**). Initially, a bias of –3 V was applied to the sample to stabilize the HRS state and downward $P^-$. After applying a positive bias, the resistivity of the film was greatly reduced, and an isothermal hysteresis was observed with a rectification property and a counterclockwise-clockwise feature enhanced by a higher $T_C$ of the FE-MgZnO. When a negative bias is applied, electrons pass over the CoPt/MgZnO SC, and the current density does not depend much on the potential barrier height nor on the direction of polarization when the applied bias is higher than the flat band potential. The *ER* value was about 430 % at +2 V which is higher than that of stack-A with an Mg content of 8% because of the increase of MgZnO bandgap and the electric polarization *P*(MgZnO). This resistive switching behavior with large hysteresis at positive voltage indicates the change in SBH at the CoPt/MgZnO interface due to the polarization reversal of MgZnO by applying an electric field that sets the resistive switching loop.

Since the semiconducting ZnO sandwiched between $Co_{0.30}Pt_{0.70}$ and Cu is thick enough to avoid direct tunneling,[50] we can assume that the junction forms a BBSD with SBHs of $\Phi_{B1}$ and $\Phi_{B2}$ the SBH at Cu and $Co_{0.30}Pt_{0.70}$ interfaces, respectively. When $V_A$ is applied, one of the diodes is reversed-biased, while the other one is forward-biased. The total current will be limited by smaller current in the reverse biased diode. Here, $R_S$ is the series resistance resulting from the two MgZnO SCs and MgZnO thin film. $R_S$ value can be determined from the linear part of the I-V curve at high negative bias, i.e., $R_S \sim \Delta V_A/\Delta I$.[76] Actually, the voltage drop $IR_S$ needs be excluded from $V_A$ to obtain the effective voltage drop across the two diodes. For a low turn-on voltage of a Schottky junction, it can be assumed that most of the voltage is distributed to the reversed-bias junctions. Finally, considering that the carrier transport across the junctions is dominated by the thermionic transport theory, and can be expressed by the following equations.[77]

$$I \approx \begin{cases} I_{S2} \exp\left(-\frac{q(V_A-IR_S)}{n_2 k_B T}\right)\left[\exp\left(\frac{q(V_A-IR_S)}{k_B T}\right)-1\right] & if\ V_A > 0 \\ I_{S1} \exp\left(+\frac{q(V_A-IR_S)}{n_1 k_B T}\right)\left[1-\exp\left(-\frac{q(V_A-IR_S)}{k_B T}\right)\right] & if\ V_A < 0 \end{cases} \quad (1)$$

$I_{S1(2)}$ is the reverse saturation current, $n_{1,2}$ is the ideality factor, $T$ is the temperature in Kelvin, $q$ is the elementary charge and $k_B$ is the Boltzmann constant. Numbers of 1 and 2 in the subscript of $I_S$, $\Phi_B$ and $n$ refer to the Cu and $Co_{0.30}Pt_{0.70}$ interfaces, respectively. Therefore, **Eq. (1)** can be rewritten in the following forms.

$$\ln\left\{\frac{I}{\exp\left(\frac{q(V_A-IR_S)}{k_B T}\right)-1}\right\} = \ln(I_{S2}) - \frac{q(V_A-IR_S)}{n_2 k_B T} \qquad for\ V_A > 0 \qquad (2)$$

$$\ln\left\{\frac{I}{1-\exp\left(-\frac{q(V_A-IR_S)}{k_B T}\right)}\right\} = \ln(I_{S1}) + \frac{q(V_A-IR_S)}{n_1 k_B T} \qquad for\ V_A < 0 \qquad (3)$$

Plotting $\ln\{I/[exp(q(V_A - IR_S)/k_BT) - 1]\}$ versus ($V_A$ – $IR_S$) for positive $V_A$ results in a linear graph which the slope and y-axis intercept directly gives $n_2$ and $I_{S2}$, respectively. From the calculated $I_{S2}$, the SBH $\Phi_{B2}$ can be calculated:

$$q\,\Phi_{B1(2)} = k_B T\,ln(S_{1(2)}\,A^*\,T^2/I_{S1(2)}) \qquad (4)$$

$S_{1(2)}$ is the SC area (circular of a diameter ϕ of 200 µm) and A* is the MgZnO effective Richardson constant (estimated as A* = 38 A cm$^{-2}$ K$^{-2}$ assuming m$_*$ = (0.32 ± 0.03) m$_0$).[45] In the same way, the parameters of the other junction can be determined by utilizing **Eq. (3)** and **(4)**. The plotting of equations **(1)** and **(2)** for both LRS and HRS states and the corresponding fitting parameters are illustrated in **Figure 4 (c)** and **(d)**.

The extracted SBHs $\Phi_{B1}$ and $\Phi_{B2}$ are summarized in **table I**. The values obtained for CoPt/MgZnO SC are consistent with the previous experimental reports of Pt/ZnO and Pt/MgZnO SCs that ranged within 390–930 meV[45,78] and a limit of 950 meV from first principles calculations for Zn-terminated ZnO interface.[79] The Cu/MgZnO SBHs are bigger than the recently reported Cu/ZnO SBHs ranged within 530–700 meV [80,81] due to the increase in the ZnO band gap by Mg doping. While there is no report about the estimation of SBH at Co$_{0.30}$Pt$_{0.70}$/(Mg,Zn)O interface, the extracted $\Phi_{B2}$ value is bigger than the effective tunneling barrier height of 300 meV from our previous reports[29,30] due to the suppression of the quantum tunneling in MgZnO-SCs. Therefore, the value of SBH at Co$_{0.30}$Pt$_{0.70}$/MgZnO interface is a consequence of the optimized MgZnO deposition by MBE which resulted in a high-quality interface as discussed in **section 3)**. Here, the ideality factors n$_{1,2}$ of MgZnO was around 1, indicating that thermionic emission is the dominant transport phenomenon, as preliminarily assumed in **Eq. (1) to (3)**. R$_S$ values are almost similar for both LRS and HRS and consistent with the thick MgZnO SC used in this report.

## 6) Ferroelectricity induced modulation of Schottky barrier heights

Accounting ferroelectric polarization reversal of $Mg_{0.2}Zn_{0.80}O$, we consider the potential barrier at the CoPt/MgZnO interface is given by $\Phi_B^\pm = \Phi \mp \Delta\Phi/2$, where $\Phi_B^\pm$ is the SBH at the $P^\pm$ state, and $\Delta\Phi$ is the SBH difference due to the MgZnO polarization reversal. In an isothermal process, after applying a positive bias, the P⁺ state is formed and the SBH at the CoPt/MgZnO interface decreases ($\Phi_B^+ = \Phi - \Delta\Phi/2$). After applying a negative bias, MgZnO is reversed to form the $P^-$ state, and the SBH increases as $\Phi_B^- = \Phi + \Delta\Phi/2$. According to the numerical fitting of **table I**, we obtained a ΔΦ(CoPt) of 26.5 meV while ΔΦ(Cu) is smaller as 5.4 meV. Another estimation of CoPt/MgZnO SBHs for both polarization states at positive biases using the thermionic injection model was performed for comparison (see **Figure S3**), which gave a ΔΦ(CoPt) in the range of 47.7–50.7 meV. These values are thought to be overestimated since only one diode contribution was considered and the fittings were poor. This confirms the validity of the adopted fitting model described in **Eqs. (2)** and **(3)**.

Concerning the CoPt/MgZnO interface, the interface charge change (*Δn*) induced by the *P* reversal was estimated by a first-principles density functional theory calculations of the structure: vacuum/Pt(3)/Co(1)/O(1)/Zn-O(8)/vacuum (atomic monolayers or bilayers), combined with scalar and fully relativistic ultrasoft pseudopotentials (USPPs) and a plane-wave basis with the generalized gradient approximation as detailed explained in our previous report.[52] The calculations incorporate a general spin-orbit coupling (SOC) with a set of natural multi-orbitals, not a simplified SOC limited to the Rashba-type under a minimal orbital set. The change of number of electrons *Δn* at the interface Co atom of $Pt_3Co$/ZnO structure is +0.02, where the P⁺ (P⁻) state corresponds to an electron depletion from (doping to) Pt-Co-O interface atoms. Although *Δn* is close in value to an Fe/MgO interface under an electric field of 0.5–1.0 V/nm, the obtained magnetic anisotropy energy (MAE) change of −2.60 erg/cm² in CoPt/ZnO is much larger than Fe/MgO, which ranges from −0.1 to −0.2 erg/cm².[82–84] This large *Δ*MAE/*Δ*n was explained by the control of *3d−5d* orbital hybridization of Co/Pt as detailly elaborated in our previous report.[52] These two facts of orbital hybridization and increase of electron at the CoPt/MgZnO interface due to *P* reversal are in agreement with the positive ΔΦ(CoPt), and hence the polarization reversal-induced SBH modulation. The experimental value of ΔΦ(CoPt) of 26.5 meV at RT is reasonable if compared to the value of 110 meV[17] for a $Co_3Pt$/ZnO or 230 meV[85] Pt/ZnO interfaces derived from first-principles calculation assuming perfect interfaces, a full polarization reversal and an absolute zero temperature, or to the other FE materials such as Pt/PZT interface with a ΔΦ(Pt) of 60 meV at RT ($P_{PZT}$ = 70 µC/cm$^2$ vs. $P_{MgZnO}$ = 8.25 µC/cm$^2$).[2,86] On the other hand, the SBH change at the Cu/MgZnO (5.45 meV) is small enough to be associated to a FE effect, and can be counted as an error of the extracted parameter *ln($I_{S1}$)* in **figure 4 (c)**. This might be explained by the formation of an inhomogeneous and Mg poor (Mg,Zn)O area at the Cu interface which weakly responds to polarization reversal. The vertical microfabrication process should not be ruled out since the top Cu is much more affected than the bottom CoPt buffer. The obtained *I-V* characteristics are similar to those of a ferroelectric semiconductor junction device.[9,87] It is necessary to stress that SBH values cannot be attributed to the formation of CuO or $Cu_2O$ interlayers at Cu/MgZnO interfaces, since the Gibbs free energy of formation for ZnO is more negative than that of CuO and $Cu_2O$ ($\Delta G_{ZnO}$ = −320.52 kJ mol$^{−1}$, $\Delta G_{CuO}$ = −129.7 kJ mol$^{−1}$, $\Delta G_{Cu_2O}$ = −149 kJ mol$^{−1}$).[88]

From the extracted ΔΦ, we estimate the *ER* ratio using the thermionic injection model derived relationship: ER = exp(ΔΦ/$k_B$T) while $k_B$T = 26 meV for 300 K. ΔΦ values of CoPt/MgZnO and Cu/MgZnO interfaces gave an ER of 280% and 123%, respectively, which is smaller than the experimental maximum ER ratio at +2 V of 430%. The other estimation using the thermionic injection model gave an ER within 624-703% which reflects an increased *P* reversal and the dominance of the CoPt/MgZnO interface. Furthermore, we calculated the screening charge density $Q_S$ and reversed polarization $P_r$ using the obtained barrier height shift $\Delta\Phi$. The relationship between $P_r$ at the metal/FE interface and the screening charges density $Q_s$ can be represented by Thomas-Fermi screening as reported by D. Pantel and M. Alexe et al.[89]

$$\Delta\Phi/2 = el_{Pt}Q_s/\varepsilon_0\varepsilon_{Pt} \quad (5); \qquad Q_s = P_r d / \left(\varepsilon_{stat}\left(\frac{l_{Pt}}{\varepsilon_{Pt}} + \frac{l_{Cu}}{\varepsilon_{Cu}}\right) + d\right) \quad (6)$$

where, $l_{Pt}$ and $l_{Cu}$ are finite screening length in the metal electrodes Pt and Cu, respectively. $\varepsilon_{Pt}$ and $\varepsilon_{Cu}$ are the ionic permittivity of the electrode materials. The finite screening length in Pt $l_{Pt}$ and ionic permittivity of Pt $\varepsilon_{Pt}$ are given by 0.5 Å and 8, respectively.[90] In the same way, $l_{Cu}$ and $\varepsilon_{Cu}$ are given by 0.55 Å and 5. The static permittivity $\varepsilon_{stat}$ of ferroelectric material $Mg_{0.2}Zn_{0.8}O$ was obtained by impedance measurements and estimated to be 17 and $d$ the film thickness of $Mg_{0.2}Zn_{0.8}O$ is 220 nm. The magnitudes of $P_r$ at both RT and 4 K are summarized in **Table II**, using the derived power-law discussed in **Section 4)**. Although the experimental ER ratio is small in the case of the simplified fitting model, the derived $P_r$ value of $Mg_{0.20}Zn_{0.80}O$ was 8.25 μC/cm$^2$ (versus P(ZnO) = 5.3 μC/cm$^2$ from Ref [[91]]) which gives $\Delta P_{sp}/x$ ratio of 147.5 mC/m$^2$. This rate was very close only to that reported by Stölzel et al.[85] of 151 ± 15 mC/m$^2$ from the experimental determination of the spontaneous polarization of pseudoferroelectric wz-MgZnO by the examination of the recombination dynamics of polar ZnO/MgZnO quantum wells (QWs) combined with a wave-functions self-consistently solver of Schrödinger and Poisson equation. For the thermionic injection models (1) and (2), $P_r$ and $\Delta P_{sp}/x$ values weren't reasonable. Additionally, the obtained $P_r$ is larger than that in our previous report (3 μC/cm$^2$) based on a tunneling junction of $Mg_{0.23}Zn_{0.77}$ barrier at 2 K,[17] which reflects the increase of the reversed MgZnO polarization domains. A Similar approach was conducted in our previous paper to correlate the experimental findings and theoretical calculations of polar ZnO-based devices.[20] The above-mentioned results indicate that we have succeeded to fabricate a ferroelectric-like $Mg_{0.20}Zn_{0.80}O$ film which is electrically polarized as large as the $Mg_{0.2}Zn_{0.80}O$ bulk on $Co_{0.30}Pt_{0.70}$ FM metal, as well as the change of the SBH with a comparable order to the case of other FE materials. Since the resistive switching effect of the CoPt/MgZnO hetero junctions dominantly occurs at the interface regardless of the origin, we expect that the CoPt/thin-MgZnO/CoPt junction could satisfy the necessary conditions of a non-volatile multi-state memory.

## 7) Conclusions

High-quality and single-crystal $Co_{0.3}Pt_{0.7}$ ferromagnet/$Mg_{0.2}Zn_{0.8}O$ hetero-junctions by MBE technique without deterioration of the magnetic properties were developed and fabricated. The resistive switching was examined with an EFC method which revealed a ferroelectricity-like $T_C$ of MgZnO close to the bulk value. A resistive switching was isothermally demonstrated at the CoPt/$Mg_{0.20}Zn_{0.80}O$ with a large hysteresis at positive bias associated with the CoPt/MgZnO SC. Parameters Fitting revealed that the SBH changed ($\Delta\Phi$) by 26.5 meV due to the reversal of MgZnO polarization. From the $\Delta\Phi$, we estimated the magnitude of reversed polarization as 3.0 (8.5) $\mu C/cm^2$ at 300 K (2 K). These results agreed with other FE-MgZnO theoretical and experimental results and proved the potential of the CoPt/MgZnO Schottky interface in the demonstration of FM/FE hybrid junctions.                    (total = 4950 words)

Table I: The extracted parameters of MgZnO-based BBSD using Eq. (2) and (3).

| | | Co$_{0.30}$Pt$_{0.70}$/MgZnO interface | | | Cu/MgZnO interface | | |
|---|---|---|---|---|---|---|---|
| | | $\Phi_{B1}$(meV) | ideality factor $n_1$ | $R_S$ (MΩ) | $\Phi_{B2}$(meV) | ideality factor $n_2$ | $R_S$ (MΩ) |
| $P_{MZO}$ States | $P^+$ (LRS) | 797.09 | 1.08 | 11.84 | 778.30 | 0.95 | 11.84 |
| | $P^-$ (HRS) | 823.60 | 1.07 | 11.56 | 783.75 | 1.02 | 11.56 |

Table II: The extracted parameters of CoPt/MgZnO interface (ER and $P_r$) in comparison of the thermionic emission model (see **Figure S3**)

| | $\Delta\Phi$ (meV) Estimation | ER analysis | | $P_r$ analysis | | |
|---|---|---|---|---|---|---|
| | | Calculated $ER_{cal}$ (%) | $ER_{cal}/ER_{exp}$ | $P_r$ (µC/cm²) @ 300 K | $P_r$ (µC/cm²) @ 2 K | $\Delta P_r$ / x (mC/m²) |
| Simplified BBSD | 26.5 | 280 | 0.7 | 1.98 | 8.25 | 147.5 |
| thermionic injection model (1) | 47.6 | 624 | 1.5 | 3.55 | 14.79 | 474.5 |
| thermionic injection model (2) | 50.7 | 703 | 1.8 | 3.79 | 15.79 | 524.5 |

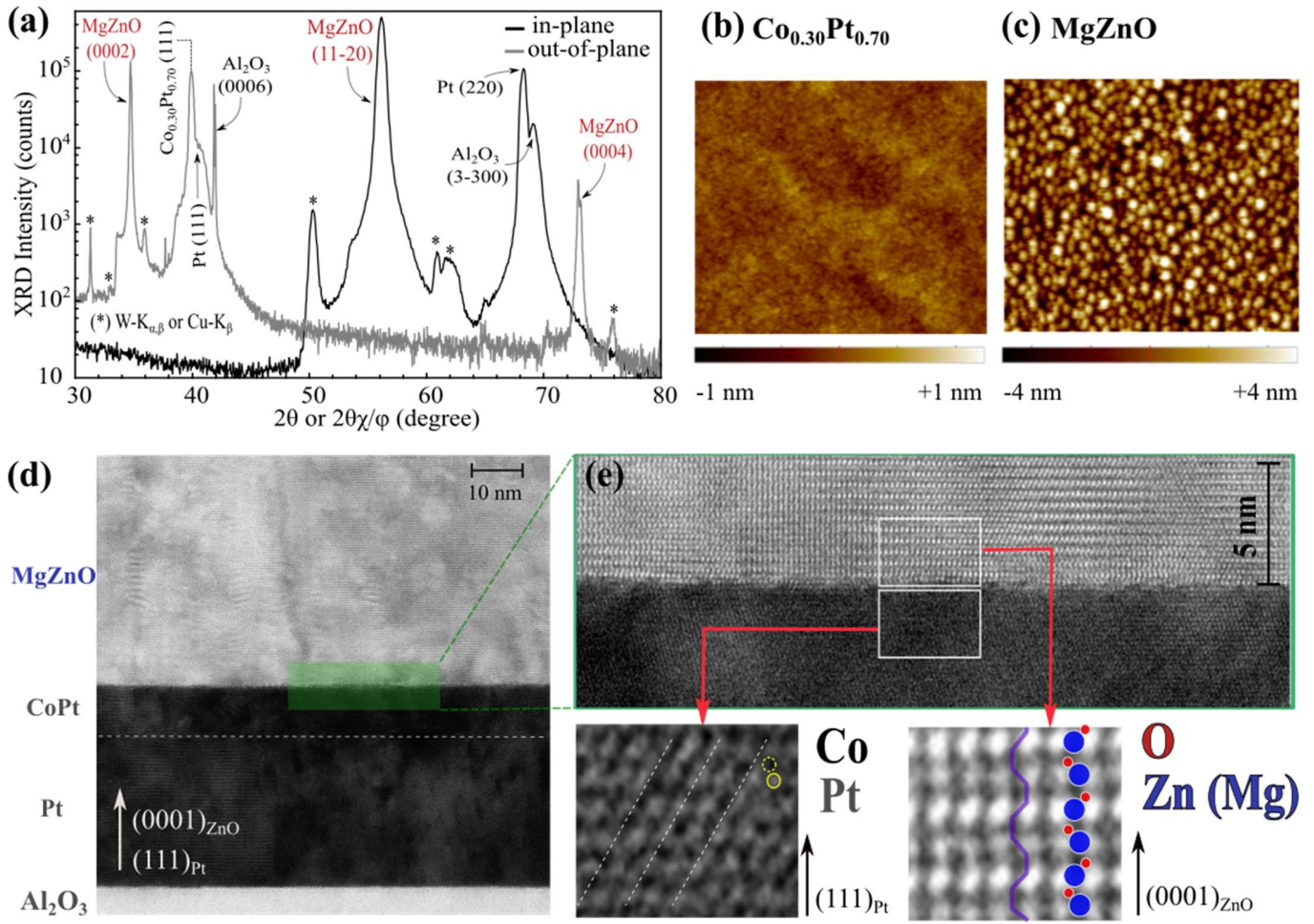

**Figure 1:** (a) the in-plane and out-of-plane XRD spectra of stack-A showing several diffraction peaks of the main materials constituting the Pt/CoPt/MgZnO stack. (b) and (c) are the 1 × 1μm² AFM scans of the Co$_{0.30}$Pt$_{0.70}$ buffer and MgZnO epilayer, respectively, for stack-A after thermal annealing. (d) and (e) are the high-resolution transmission electron microscopy (TEM) images of Al$_2$O$_3$/Pt/Co$_{0.30}$Pt$_{0.70}$/MgZnO (Mg = 20%) stack along [0001] axis and an enlarged region at the Co$_{0.30}$Pt$_{0.70}$/MgZnO interface as indicated by the green rectangle. The bottom part of (e) shows the *wz*-ZnO(0001) zigzag and the fcc-Co$_{0.30}$Pt$_{0.70}$ (111) atomic alignment.

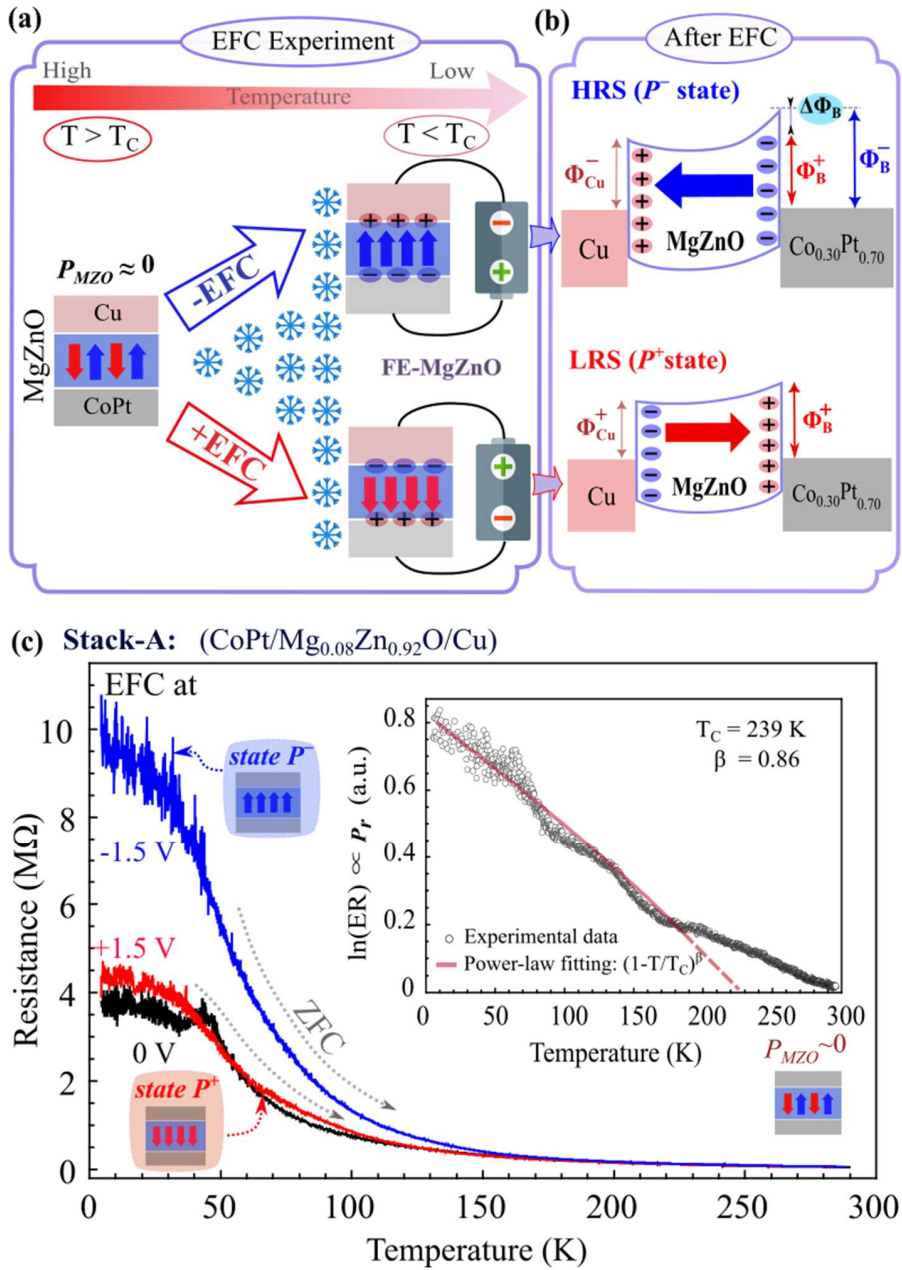

**Figure 2:** (a) A schematic of the CoPt/MgZnO stack and the electric-field cooling (EFC) procedure from a temperature higher than the $T_C$ of FE-MgZnO (300K) for Mg = 8 %. At the lowest temperature, the reversal of MgZnO electric polarization ($P$) to $P^+$ and $P^-$ states after ±EFC changes the surface charge at the CoPt/MgZnO interface. (b) the CoPt/MgZnO/Cu diagram during the zero-field heating (ZFH) gave a high (low) device resistance and SBH of $\Phi_B^-$ ($\Phi_B^+$) after the -EFC (+EFC) process. The SBH change $\Delta\Phi_B$ was also displayed. As summarized in table I, the SBH change is prominent at the CoPt/ $Mg_{0.20}Zn_{0.80}O$ interface. (c) the temperature dependence of the ZFH resistance after 0 V, +1.5 V and −1.5 V EFC of a stack-A pillar measured at +10 mV. Inset: temperature dependence of the $ln$(ER) and a fitting with the power-law, which reflects the ferroelectricity of MgZnO thin film similar to results of Refs [17].

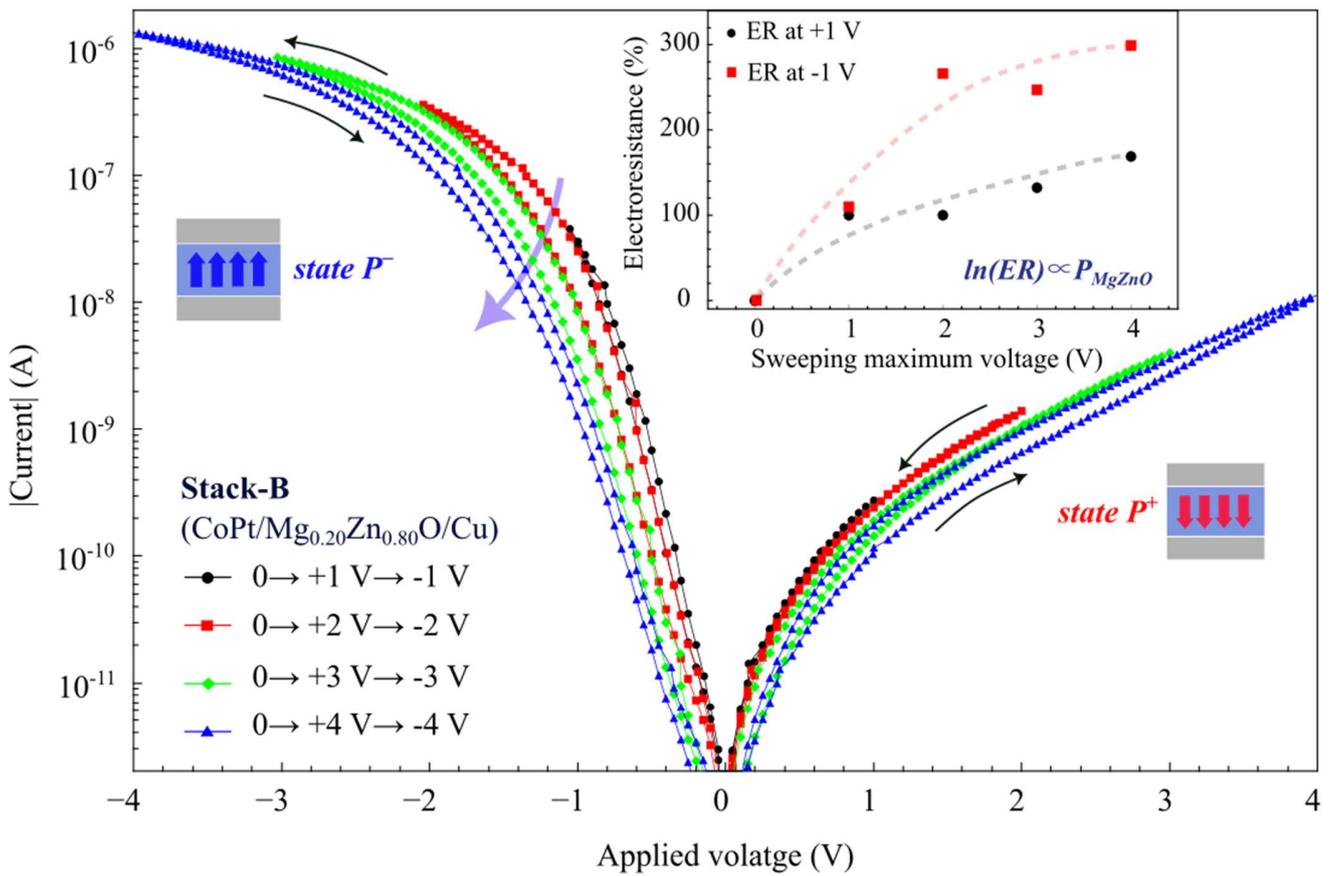

**Figure 3:** Current-voltage (*I-V*) characteristics of a CPP pilar of stack-B upon the application of consecutive DC double-sweep voltage ramp as 0→ +$V_{max}$ → –$V_{max}$ → 0 V, where $V_{max}$ = 1, 2, 3, 4 V. The black arrows show the ramping sequence, and the big violet arrow is an eye guide for the resistance increase by the ferroelectric resistive switching. The inset is the sweeping maximum voltage ($V_{max}$) dependence of the ER ratio calculated at ±1 V. The saturation tendency of ER ratio is visualized by the dotted lines.

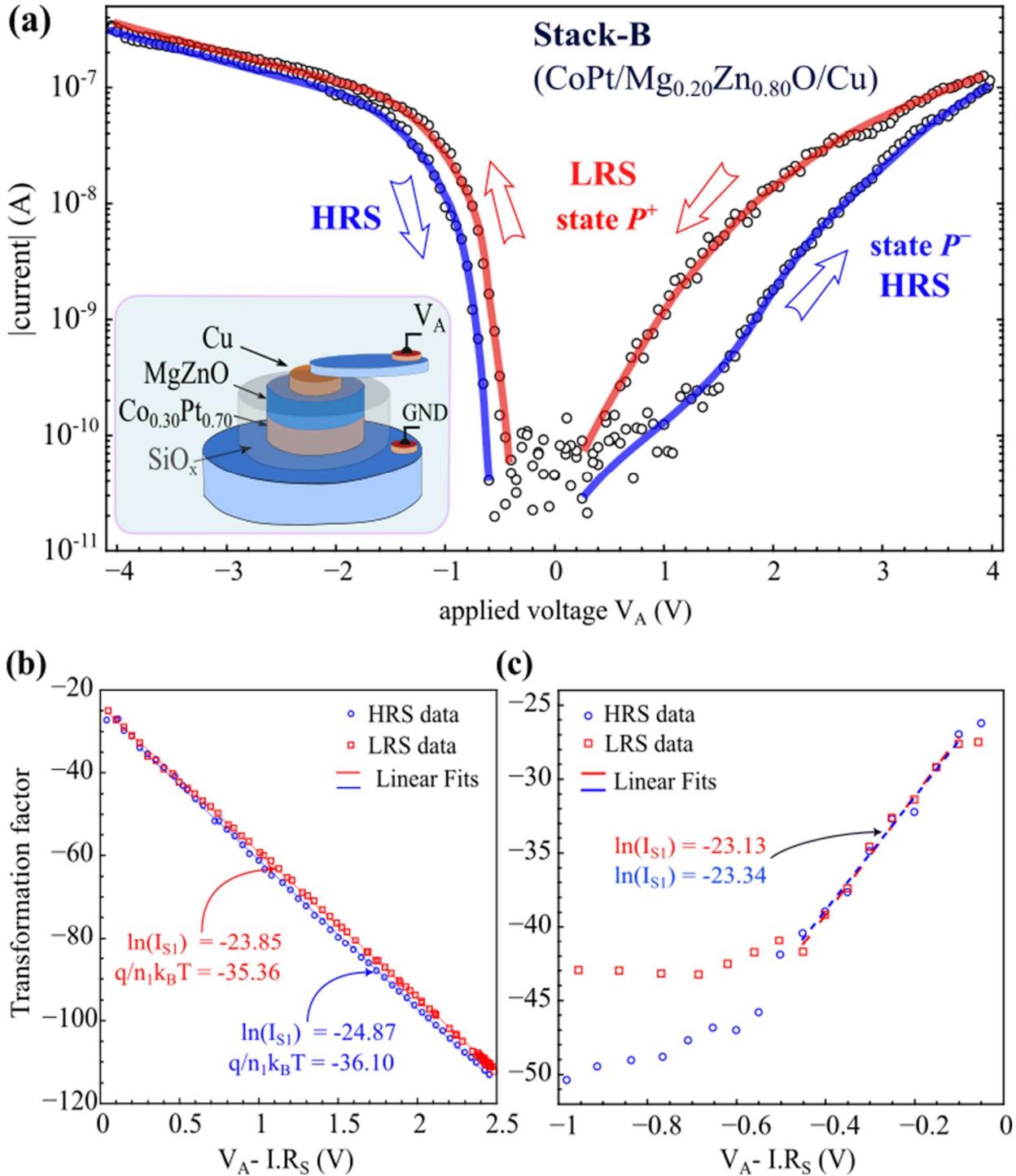

**Figure 4: (a)** Current-voltage (*I-V*) characteristics of a CPP pilar of stack-B for a DC voltage loop ranging from 0 → +4 V → −4 V → 0 V on 200 μm-ϕ Cu top electrode. −3 V was applied on this sample before the measurement for 10-20 seconds to generate the HRS instead of the EFC process. The inset is a schematic presentation of the CPP device after microfabrication. (b) and (c) are the I-V fittings for positive and negative $V_A$, respectively, for both $P^+$ and $P^-$ states. The obtained parameters are mentioned inside the graph and summarized in table I.

## Authors contributions:

M.B: Conceptualization, Methodology, Investigation, Formal analysis, Funding acquisition, Writing - Original Draft and Writing – review & editing.

M. A. M: Conceptualization, Methodology, Investigation and Data curation.

G.M Jr: Formal analysis, Data curation and Writing - Original Draft.

T.N: Methodology, Formal analysis and Investigation.

S. M: Methodology and Funding acquisition.

C. C and W.K.P: contributed to the Formal analysis, and Writing – review & editing.

M. B: mohamed.belmoubarik@um6p.ma ;            0000-0003-3592-1259
M. A. M: mahdawi@tohoku.ac.jp;                 0000-0001-7021-0829
G. M Jr:   george.junior@inl.int;              0000-0002-2795-2111
T. N: nozaki.tomohiro@aist.go.jp;              0000-0002-6220-7344
C. M: claudia.aa.coelho@gmail.com;             0000-0003-0730-5651
M. S:   sahashi@ecei.tohoku.ac.jp;             non
W. K. P: pengwengkung@sslab.org.cn;            0000-0002-7984-9319

## Data Availability

The data that support the findings of this study are available from the corresponding author upon reasonable request.

## Declaration of Competing Interest

The authors declare that they have no known competing financial interests or personal relationships that could have appeared to influence the work reported in this paper.

## Acknowledgment

This work was sponsored by Japan Society for the Promotion of Science (Grant No. 25-5806).